\documentclass[final]{aipproc}

\layoutstyle{8x11double}

\usepackage{graphicx}
\usepackage{txfonts}
\usepackage{natbib}
\input epsf.sty

\newcommand{\Mpc}{$h^{-1}$\thinspace Mpc}

\def\aj{AJ}%
\def\araa{ARA\&A}%
\def\apj{ApJ}%
\def\apjl{ApJ}%
\def\aap{A\&A}%
\def\azh{AZh}%
\def\mnras{MNRAS}%
\def\nat{Nature}%
\begin{document}

\sloppy

\title{Large scale structure of the Universe}


\author {Jaan Einasto}
{address={Tartu Observatory, EE-61602 T\~oravere, Estonia}
}
\begin{abstract}
  A short overview is given on the development of our present paradigm of the
  large scale structure of the Universe with emphasis on the role of
  Ya. B. Zeldovich. Next we use the Sloan Digital Sky Survey data and show
  that the distribution of phases of density waves of various scale in the
  present-day Universe are correlated.  Using numerical simulations of
  structure evolution we show that the skeleton of the cosmic web was present
  already in an early stage of the evolution of structure. The positions of
  maxima and minima of density waves (their phases) are the more stable, the
  larger is the wavelength.  The birth of the first generation of stars
  occured most probably in the central regions of rich proto-superclusters
  where the density was highest in the early Universe.
\end{abstract}

\maketitle

\section{Introduction: Ya. B. Zeldovich and the development of our
  understanding of the large-scale structure of the Universe}

This conference is devoted to Ya. B. Zeldovich, to celebrate his 95th
birthday, so let me start with an overview of our contacts with
Ya. B. Zeldovich and his group, in particular of his role in the development
of the modern view of the large scale structure and the evolution of the
Universe. Thereafter I will analyse the luminosity density field of galaxies
of the Sloan Digital Sky Survey (SDSS).  Finally I will use numerical
simulations of evolution of structure to follow the changes of phases
(positions of maxima and minima) of density waves of various scales, using
wavelet techniques.

\begin{figure*}[ht]
\centering
\resizebox{0.95\textwidth}{!}
{\includegraphics{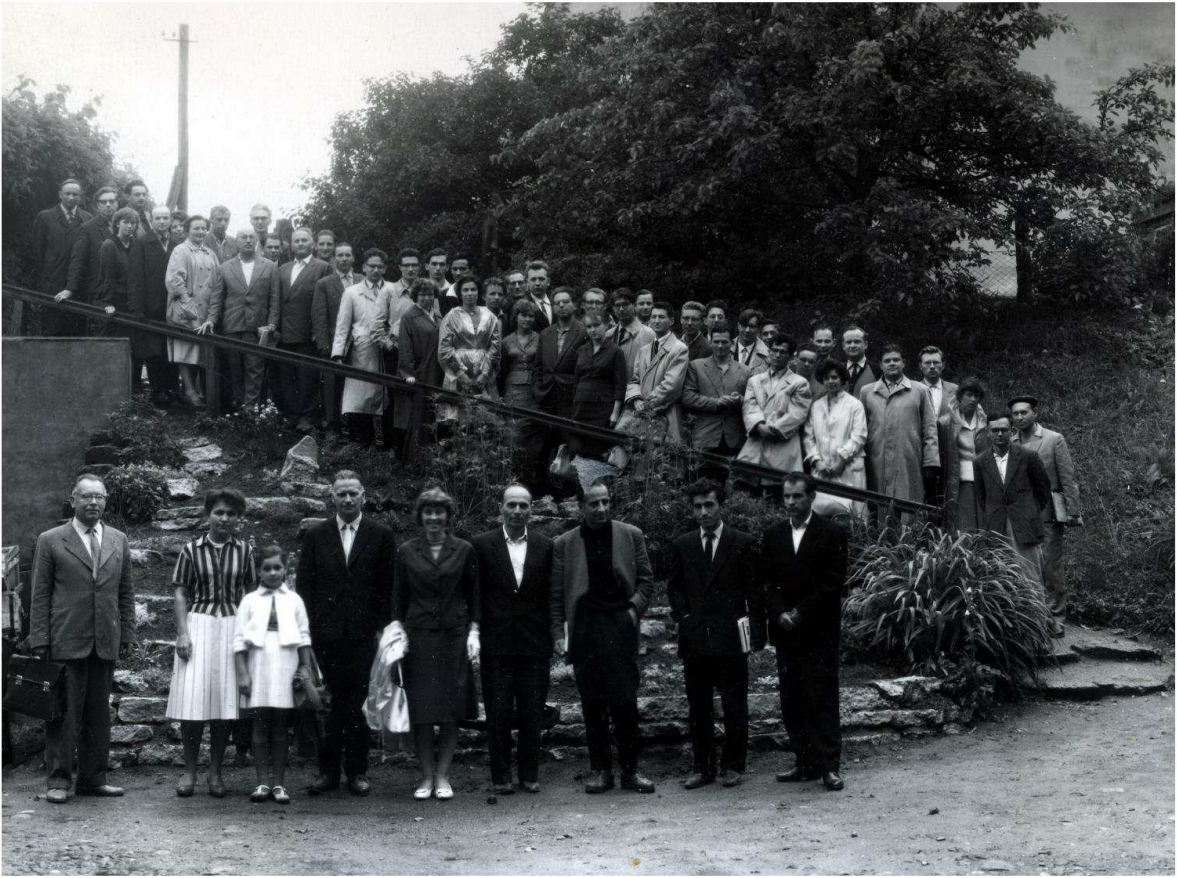}}
\caption{Participants of the Cosmology Summer School 1962 visiting old
  observatory in Tartu. }
\label{fig:zeld62}
\end{figure*}

Tartu astronomers met Ya. B. Zeldovich for the first time in 1962.
The main building of the new observatory in T\~oravere was not yet
completed, but under the initiative of Ya. B. Zeldovich a summer
school on cosmology was held there, one of the first of similar summer
and winter schools.  Fig.~\ref{fig:zeld62} shows participants of the
school visiting the old Tartu Observatory. In 1960s Tartu astronomers
participated in Astrophysics Seminars in Sternberg Institute,
organized by Ya. B. Zeldovich, during their visits to Moscow.  Here
astronomers from all Moscow astronomical institutes gathered to
discuss actual problems of astrophysics.

My own first report in this seminar was in 1971, when I made a summary of my
doctoral thesis on galactic modeling.  This report was a starting point of our
collaboration with Zeldovich and his group. Next year members of our cosmology
group were invited to participate in the Arch\~oz Winter School on
astrophysics, where I made a more detailed report of new galactic models.  In
January 1974 the School was held near the Elbrus mountain in a winter
resort. I reported our very fresh results on the study of the dynamics of
satellite galaxies. Our data suggested {\em that all giant galaxies have
  massive invisible coronas, which exceed masses of known stellar populations
  at least tenfold.  Thus dark matter must be the dominating component in the
  whole universe}.  In the discussion after the talk two questions dominated:
What is the physical nature of the dark matter?  and What is its role in the
evolution of the Universe?

The importance of dark matter for cosmological studies was evident,
thus Tartu astronomers organized in January 1975 a conference in
Tallinn devoted solely to dark matter (\citet{Doroshkevich:1975}).
Historically this was the first conference on dark matter. In addition
to dynamical and morphological properties of galaxy systems
(\citet{Einasto:1974, Einasto:1974b}) the nature of galactic coronas
was discussed.  Here are titles of some talks held at this conference:
\citet{Zeldovich:1975}: {\em ``Deuterium nucleosynthesis in the hot
  Universe and the density of matter''}; \citet{Komberg:1975a}: {\em
  ``Physical nature of galactic coronas''}; and \citet{Jaaniste:1975}:
{\em ``Properties of stellar halos''}.  These studies reported at the
conference demonstrated that both the stellar as well as gaseous
models for coronas have difficulties. It is very difficult to explain
physical properties of the stellar corona, since all old spheroidal
populations have a low mass-to-luminosity ratio whereas this ratio for
the corona is very high. Also, no fast-moving stars as possible
candidates for stellar coronas were found.  Gaseous corona cannot
consist of neutral gas (intergalactic hot gas would ionize the coronal
gas), but a corona consisting of ionized gas would be
observable. Actually, at least part of the corona must be gaseous, as
demonstrated by the morphology of satellite galaxies: elliptical
(non--gaseous) companions lie close to the primary galaxy whereas
spiral and irregular (gaseous) companions of the same luminosity have
larger distances from the primary galaxy, as shown by
\citet{Einasto:1974b}.  The amount of the hot gas is, however,
insufficient to explain flat rotation curves of galaxies.

\begin{figure}[ht]
\centering
\resizebox{0.45\textwidth}{!}
{\includegraphics{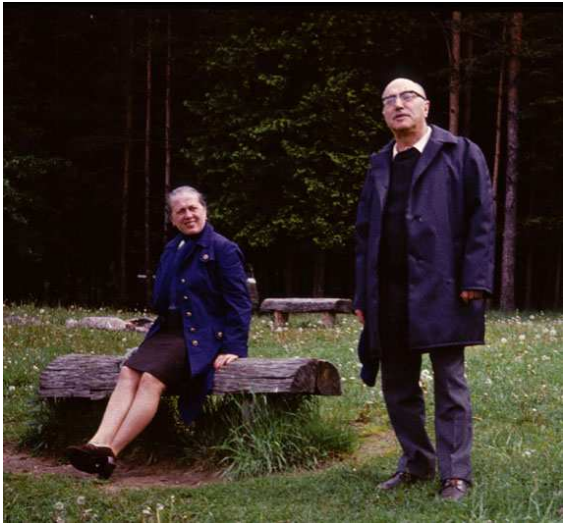}}
\caption{Zeldovich with his wife visiting Estonia. }
\label{fig:zeld78}
\end{figure}

\begin{figure}[ht]
\centering
\resizebox{0.48\textwidth}{!}
{\includegraphics{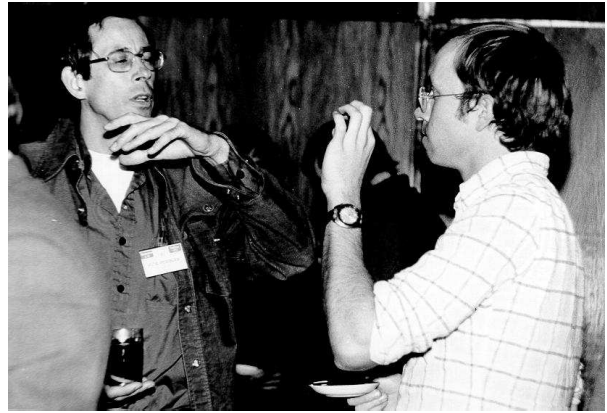}}
\caption{Jim Peebles and Scott Tremaine at the IAU Symposium on Large Scale
  Structure of the Universe in Tallinn 1977. }
\label{fig:peebles}
\end{figure}

Whatever the nature of galactic coronas, it was evident that dark matter as
the dominant population of the universe determines details of physical
processes of the formation and evolution of galaxies.  Zeldovich asked Tartu
astronomers for help in solving the question: Can we find observational
evidence that can be used to discriminate between various theories of galaxy
formation?  At this time there were two basic rivaling theories of structure
formation: the pancake theory by \citet{Zeldovich:1970}, and the hierarchical
clustering theory by \citet{Peebles:1971} (see Figures \ref{fig:zeld78} and
\ref{fig:peebles}).  According to the Zeldovich scenario the structure forms
top-down: first matter collects into pancakes and then fragments to form
smaller units.  In the hierarchical clustering scenario the order of the
formation of systems is opposite: first small-scale systems (star-cluster
sized objects) form, and by clustering systems of larger size (galaxies,
clusters of galaxies) form; this was called the bottom-up scenario.

\begin{figure}[ht]
\centering
\resizebox{0.45\textwidth}{!}
{\includegraphics{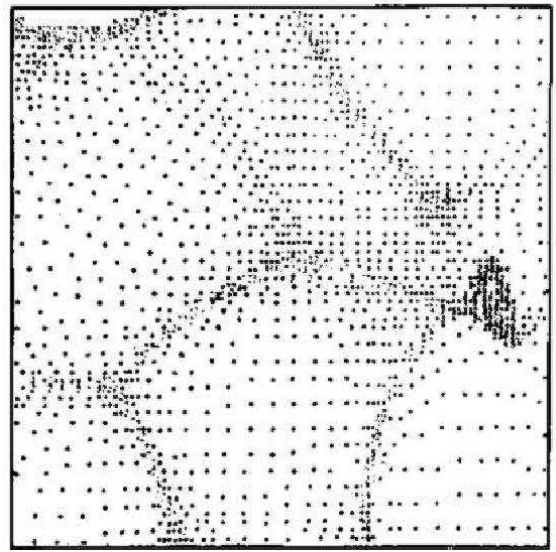}}
\caption{Distribution of particles in a N-body simulation by the Zeldovich
  group in 1975.  }
\label{fig:nbody}
\end{figure}

When solving the Zeldovich question we recalled our previous experience in the
study of galactic populations: structural properties of populations hold the
memory of their previous evolution and formation.  Random velocities of
galaxies are of the order of several hundred km/s, thus during the whole
lifetime of the Universe galaxies have moved from their place of origin only
by about 1~\Mpc\ (we use in this paper the Hubble constant in the units of
$H_0 = 100~h$ km~s$^{-1}$~Mpc$^{-1}$).  In other words -- if there exist some
regularities in the distribution of galaxies, then these regularities must
reflect the conditions in the Universe during the formation of galaxies.  What
is expected in the pancake model is shown in Fig.~\ref{fig:nbody}.  This
simulation of structure formation was made in mid-1970s in the Institute of
Applied Mathematics with the largest available computer at this time by the
Zeldovich group. In the simulation a network of high- and low-density regions
was seen: high-density regions form cells which surround large under-dense
regions.

\begin{figure*}[ht]
\centering
\resizebox{0.48\textwidth}{!}
{\includegraphics{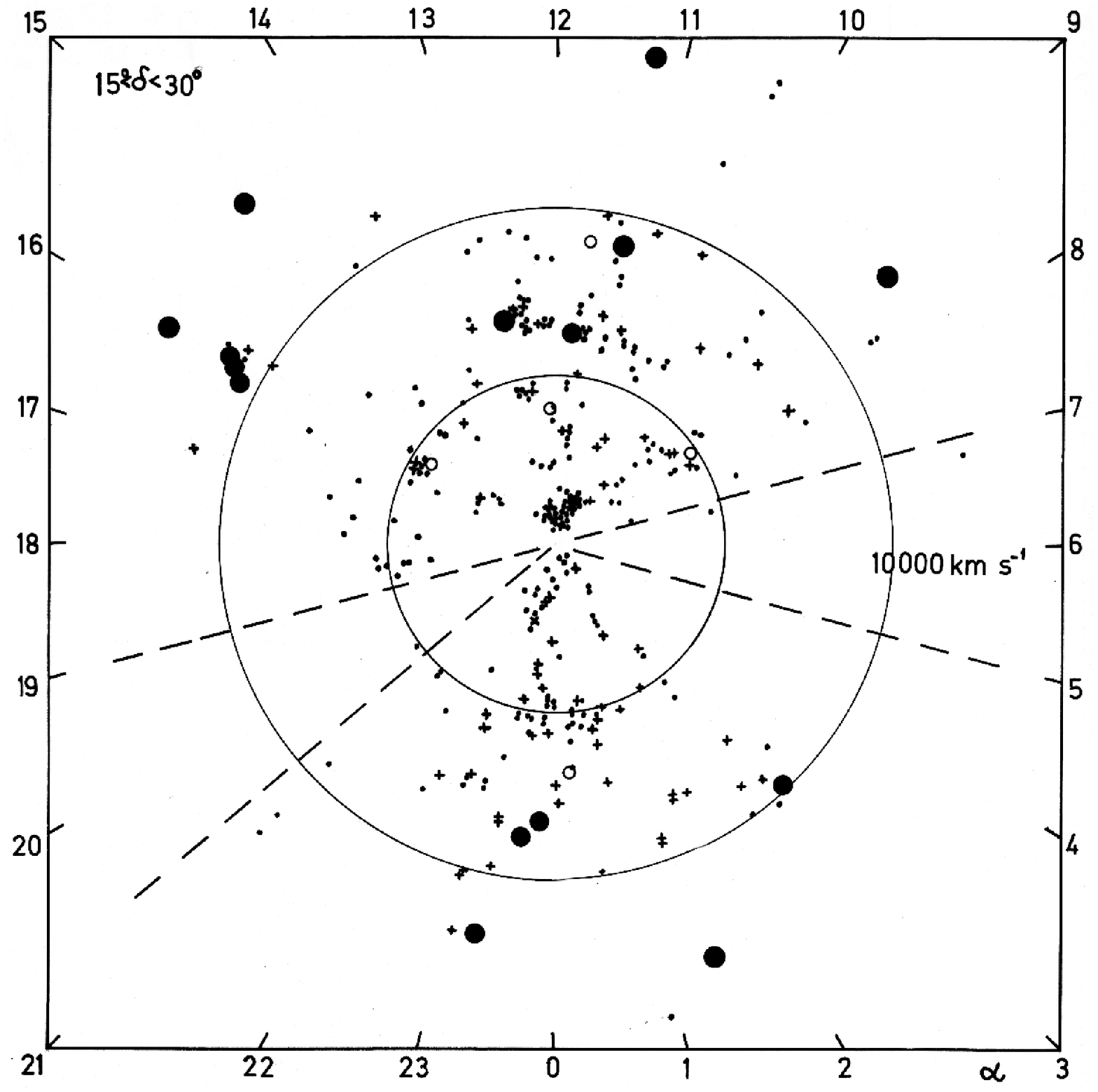}}
\resizebox{0.48\textwidth}{!}
{\includegraphics{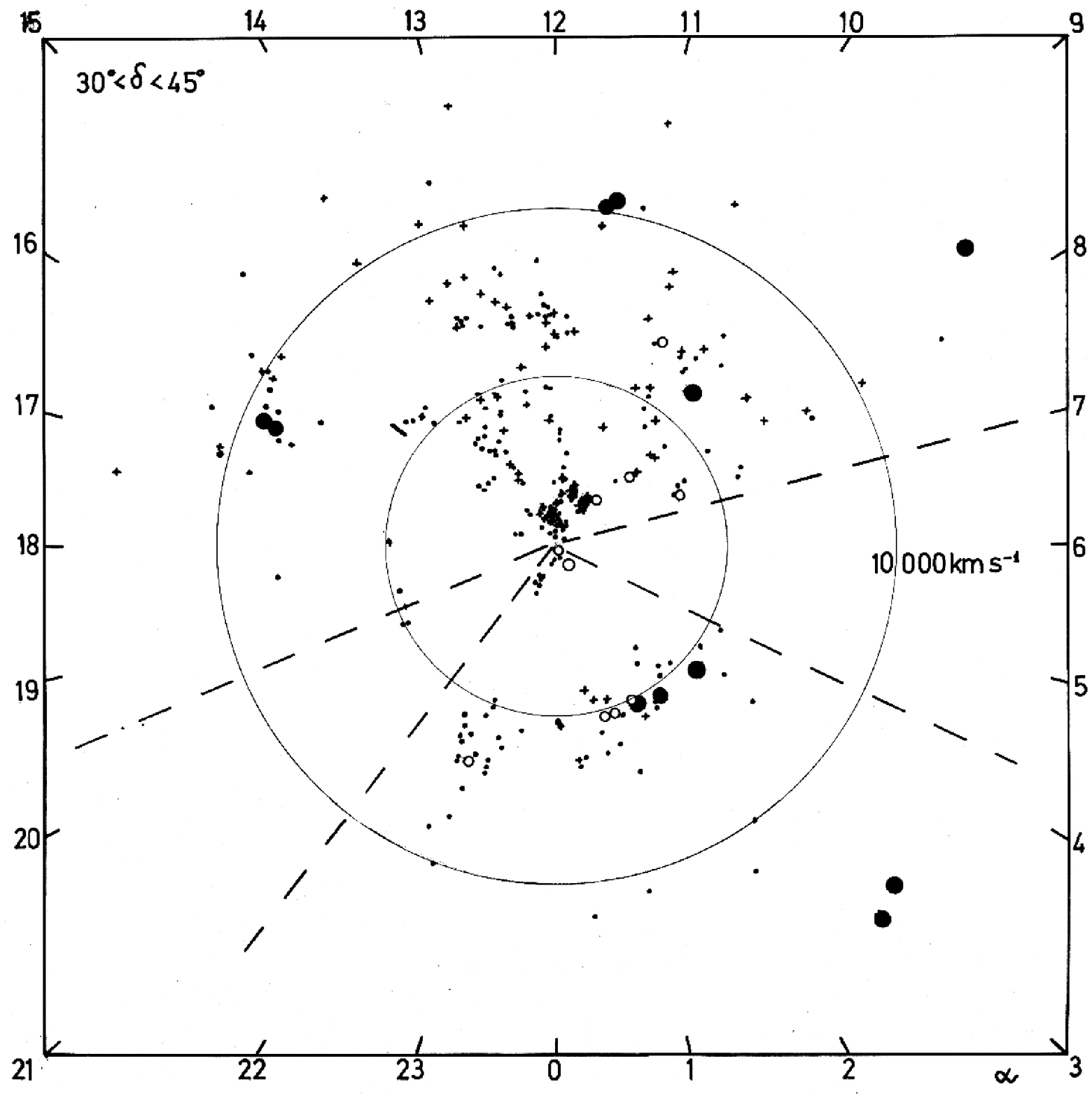}}
\caption{Wedge diagrams for two declination zones. Filled circles show rich
  clusters of galaxies, open circles -- groups, dots -- galaxies, crosses --
  Markarian galaxies. In the $15^\circ - 30^\circ$ zone two rich clusters at
  RA about 12 h are the main clusters of the Coma supercluster, in the
  $30^\circ - 45^\circ$ zone the clusters at RA about 3 h belong to the main
  chain of clusters and galaxies of the Perseus-Pisces supercluster. Note the
  complete absence of galaxies in front of the Perseus-Pisces supercluster,
  and galaxy chains leading from the Local supercluster towards the Coma
  supercluster in the upper zone and towards the Perseus-Pisces supercluster
  in the lower zone (\citet{Joeveer:1978a}).  }
\label{fig:wedges}
\end{figure*}

When we started this study, first all-sky complete redshift surveys of
galaxies were just available: the Shapley-Adams revised catalogue
(\citet{Sandage:1981a}), complete up to the magnitude 13.5, and the Second Revised
Catalogue of Galaxies (\citet{de-Vaucouleurs:1976c}), containing data on even
fainter galaxies.  The three-dimensional distribution of galaxies, groups and
clusters of galaxies can be visualised using wedge diagrams. In these
diagrams, where galaxies as well as groups and clusters of galaxies were
plotted -- see Fig.~\ref{fig:wedges}, regularity was clearly seen: {\em
  galaxies and clusters are concentrated to identical essentially
  one-dimensional systems, and the space between these systems is practically
  empty}.

The distribution was quite similar to the distribution of simulation particles
in a numerical simulation of the evolution of the structure of the Universe
prepared by the Zeldovich group, Fig.~\ref{fig:nbody}.  We had the impression
that observed high-density regions -- superclusters of galaxies -- could be
identified with Zeldovich pancakes.

We reported our results (\citet{Joeveer:1978a}) at the IAU symposium on
Large-Scale Structure of the Universe in Tallinn 1977, the first conference on
this topic. The main results were: (1) galaxies, groups and clusters of
galaxies are not randomly distributed but form chains, converging in
superclusters; (2) the space between galaxy chains contains almost no galaxies
and forms holes (voids) of diameters of $20 \dots 50$~\Mpc, as seen from the wedge
diagrams in Fig.~\ref{fig:wedges}.  The presence of holes (voids) in the
distribution of galaxies was reported also by other groups:
\citet{Tully:1978}, \citet{Tifft:1978} and \citet{Tarenghi:1978} in the Local,
Coma and Hercules superclusters, respectively.  Theoretical interpretation of
the cellular structure was discussed by \citet{Zeldovich:1978}.

The most important aspect of the structure of superclusters is the fine
structure of chains: clusters and groups within a chain are elongated along the
chain, as seen in the long chain of clusters, groups and galaxies
of the Perseus-Pisces supercluster.  This chain is located almost
perpendicular to the line of sight, and the scatters of positions of
clusters/groups along the chain in radial (redshift) and tangential directions
are practically identical.  This demonstrates that the chain is essentially an
one-dimensional structure.  A direct consequence of this observation is that
galaxies and groups/clusters of the chain are already formed within the
chain.  A later inflow from random locations to the chain is excluded, since
in this case it would be impossible to stop galaxies and clusters in the chain
after the inflow.  This is possible only in a gaseous dissipative
medium before the formation of galaxies.

New data gave strong support to the Zeldovich pancake scenario.  However, some
important differences between the model and observations were evident.  First
of all, there exists a rarefied population of simulation particles in voids,
absent in real data. This was the first indication for the presence of
physical biasing in galaxy formation -- there must be primordial gas and dark
matter in voids (because gravity cannot evacuate voids completely), but due to
low matter density no galaxy formation takes place here.

The second difference lies in the structure of galaxy systems in high-density
regions: in the original pancake model large-scale structures (superclusters)
have rather diffuse forms, real superclusters consist of multiple intertwined
filaments (\citet{Joeveer:1978a, Zeldovich:1982, Oort:1983}).  The reason of
this discrepancy was clarified several years later.

Let me recall one more historical moment in our search for understanding the
cosmic evolution in late 1970s.  At the Tallinn symposium
\citet{Parijskij:1978} made a report on his search for primordial
perturbations in the Universe with RATAN-600, the largest and most sensitive
radio telescope at this time. No temperature fluctuations were found, the
upper limit was about $10^{-4}$ of the mean temperature.  After the talk
Zeldovich discussed these results with Parijskij and expressed his doubts that
something must be wrong in his observations.  Theoretical calculations show
that at the epoch of recombination the density (and temperature) fluctuations
must have an amplitude of the order of $10^{-3}$, otherwise the structure
cannot form, since gravitational instability that is responsible for the
growth of the amplitude of fluctuations, works very slowly in an expanding
Universe.

The above arguments concern fluctuations of the baryonic gas density.
Then astronomers recalled the possible existence of non-baryonic particles,
such as heavy neutrinos. This suggestion was made independently by several
astronomers (\citet{Cowsik:1973, Szalay:1976, Tremaine:1979,
  Doroshkevich:1980b, Chernin:1981, Bond:1983b}) and others.  They found that
if dark matter consists of heavy neutrinos, then this helps to explain the
paradox of small temperature fluctuations of the cosmic microwave background
radiation.  This problem was discussed at a conference in Tallinn in April
1981.  Recent experiments by a Moscow physicist Lyubimov were announced, which
suggested that neutrinos have masses. If so, then the growth of perturbations
in a neutrino-dominated medium can start much earlier than in a baryonic
medium, and at the time of recombination perturbations may have amplitudes
large enough for structure formation.  At the conference banquet Zeldovich
gave an enthusiastic speech: {\em ``Observers work hard in sleepless nights to
  collect data; theorists interpret observations, are often in error, correct
  their errors and try again; and there are only very rare moments of
  clarification.  Today it is one of such rare moments when we have a holy
  feeling of understanding the secrets of Nature.''}  Non-baryonic dark matter
is needed to start structure formation early enough.

However, in the neutrino-dominated dark matter small-scale perturbations are
damped.  This prevents the formation of fine structure, such as
galaxy-sized halos and filaments.  Motivated by the observational problems
with neutrino dark matter, a new dark matter scenario was suggested by
\citet{Blumenthal:1982, Bond:1982, Pagels:1982, Peebles:1982, Bond:1983,
  Doroshkevich:1984} with hypothetical particles as axions, gravitinos,
photinos or unstable neutrinos playing the role of dark matter. This model was
called the Cold Dark Matter (CDM) model, in contrast to the neutrino-based Hot
Dark Matter model. 

Numerical simulations of the evolution of structure for the hot and
cold dark matter were compared by \citet{Melott:1983}, and by
\citet{White:1983, White:1987} (the standard model with the density
parameter $\Omega_m = 1$, SCDM).  In contrast to the HDM model, in the
SCDM scenario formation of structure starts early and superclusters
consist of a network of small galaxy filaments, similar to the
observed distribution of galaxies.  Thus SCDM simulations reproduce
quite well the observed structure with clusters, filaments and voids,
including its quantitative characteristics (percolation or
connectivity, the multiplicity distribution of systems of galaxies).

As a further step to improve the model, the cosmological constant, $\Lambda$,
was incorporated into the scheme. Arguments favouring a model with the
cosmological constant were suggested already by \citet{Gunn:1975, Turner:1984,
  Kofman:1985}: combined constraints on the density of the Universe, ages of
galaxies, and baryon nucleosynthesis.  Numerical models with the cosmological
$\Lambda-$term were developed by \citet{Gramann:1988}, a post-graduate student
of Enn Saar. Comparison of the SCDM and $\Lambda$CDM models shows that the
structure of the cosmic web is similar in both models.  However, in order to
get a correct amplitude of density fluctuations, the evolution of the SCDM
model has to be stopped at an earlier epoch, thus the $\Lambda$CDM model is
superior of the SCDM model. The $\Lambda$CDM model combines essential aspects
of both original structure formation models, the pancake and the hierarchical
clustering scenario.

One difficulty of the original pancake scenario by Zeldovich is the shape of
objects formed during the collapse.  It was assumed that forming systems are
flat pancake-like objects, whereas dominant features of the cosmic web are
filaments.  This discrepancy has found an explanation by \citet{Bond:1996}.
They showed that in most cases just essentially one-dimensional structures,
i.e. filaments form.  Two coauthors of this study, Kofman and Pogosyan, are
graduates of the Tartu University and former members of our cosmology group.

\section{Distribution of galaxies in the Sloan survey}

Presently the largest project to map the Universe is the Sloan Digital Sky
Survey (SDSS) (\citet{York:2000, Stoughton:2002, Zehavi:2002}).  The goal is to
map a quarter of the entire sky: to determine the positions and photometric data
in 5 spectral bands for galaxies and quasars (about 100 million objects) down
to the red magnitude {\tt r = 23}, and the redshifts of all galaxies down to {\tt
  r = 17.7} (about 1 million galaxies).  All seven data releases have been made
public. This has allowed to map the largest volume of the Universe so far.

\begin{figure}[ht]
\begin{minipage}[t]{0.48\textwidth}
\centering
\resizebox{0.980\textwidth}{!}{\includegraphics{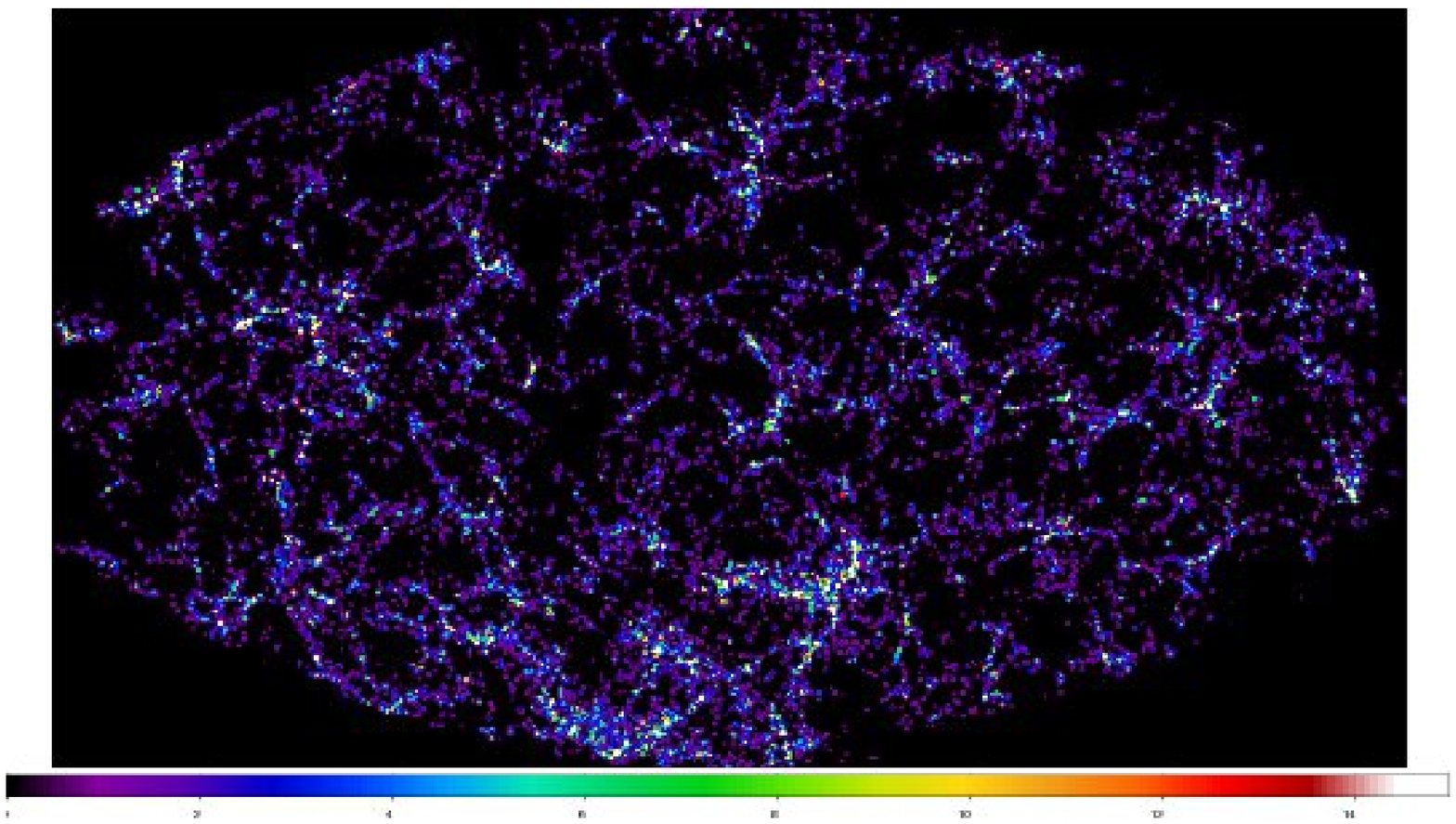}}\\
\resizebox{0.980\textwidth}{!}{\includegraphics{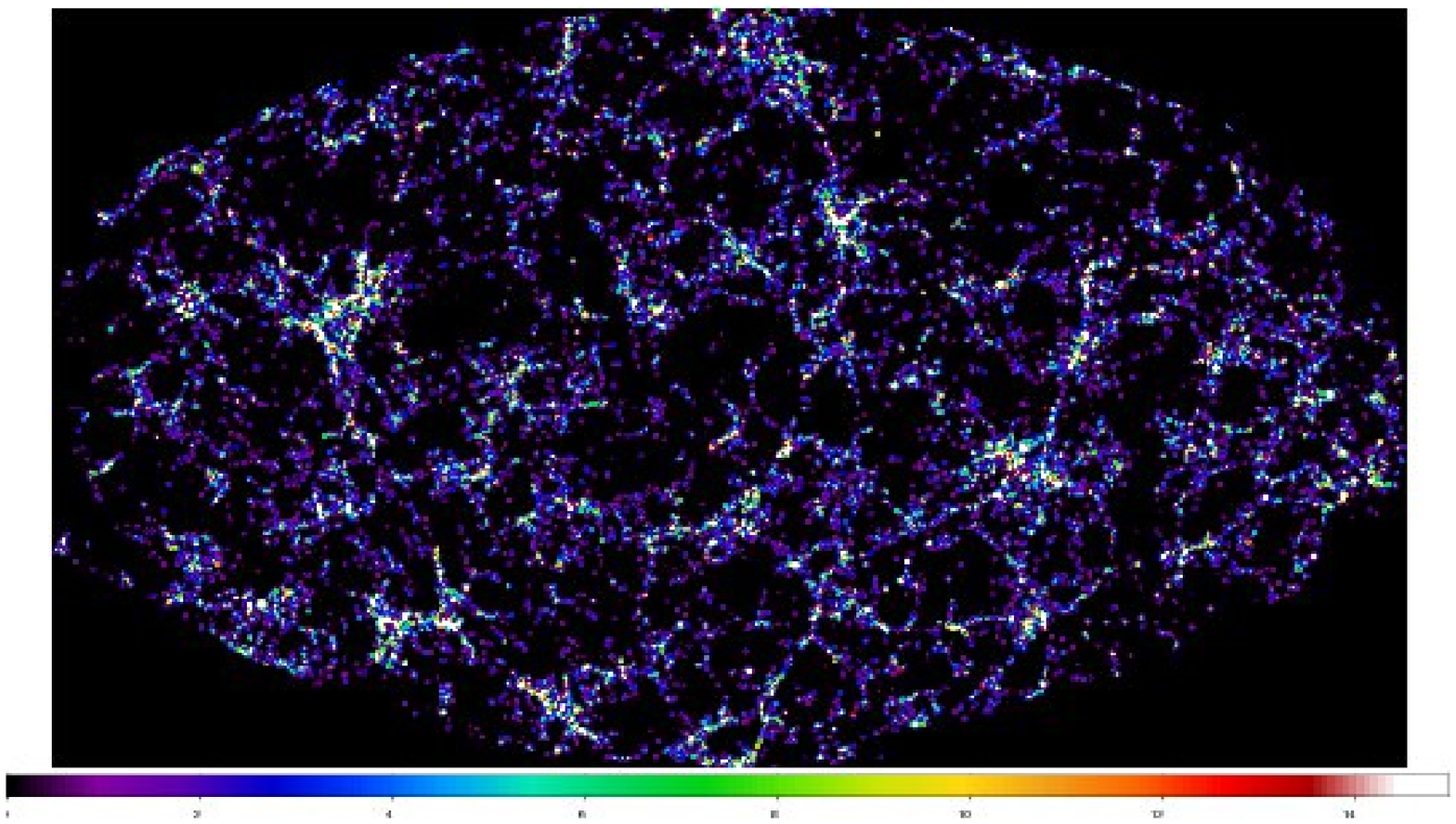}}
\end{minipage}
\caption{The luminosity density field in spherical shells of 10~\Mpc\ thickness
  and distances of 240~\Mpc\ and 325~\Mpc\ is shown in the upper and lower panels,
  respectively.  In the upper panel the very rich supercluster in the lower region
  is SCL126 (\citet{Einasto:2001}), or the Sloan Great Wall.  Cellular
  arrangement of superclusters and filaments surrounding voids is clearly
  seen. The angular diameters of cells in the 325 ~\Mpc\ shell are smaller than in
  240~\Mpc\ shell.}
\label{fig:dr7_240}
\end{figure}

The luminosity density fields of two spherical shells of the full Sloan Survey
in the Northern hemisphere are shown in Fig.~\ref{fig:dr7_240}.  The luminosity
densities were corrected for the incompleteness due to the distance dependent
lower luminosity limit of the flux-limited survey. The cosmic web with
filamentary superclusters and voids is very well seen.  Faint filaments
crossing large voids can be seen also.  The web has a characteristic cellular
pattern, the diameter of one cell -- a void surrounded by superclusters -- is
fairly constant, about 100~\Mpc.  Since shells are located at different
distances from us, the angular diameters of faraway cells are
smaller.  The great variability of supercluster richness can be easily
seen.

The variability of supercluster richness is due to joint action of density
perturbations of different scales.  To understand better the role of
perturbations of various scale we have decomposed the density field into
components of various scale using a wavelet technique.

\begin{figure*}[ht]
\begin{minipage}[h]{0.48\textwidth}
\centering
\hspace{1mm}
\resizebox{0.40\textwidth}{!}{\includegraphics*{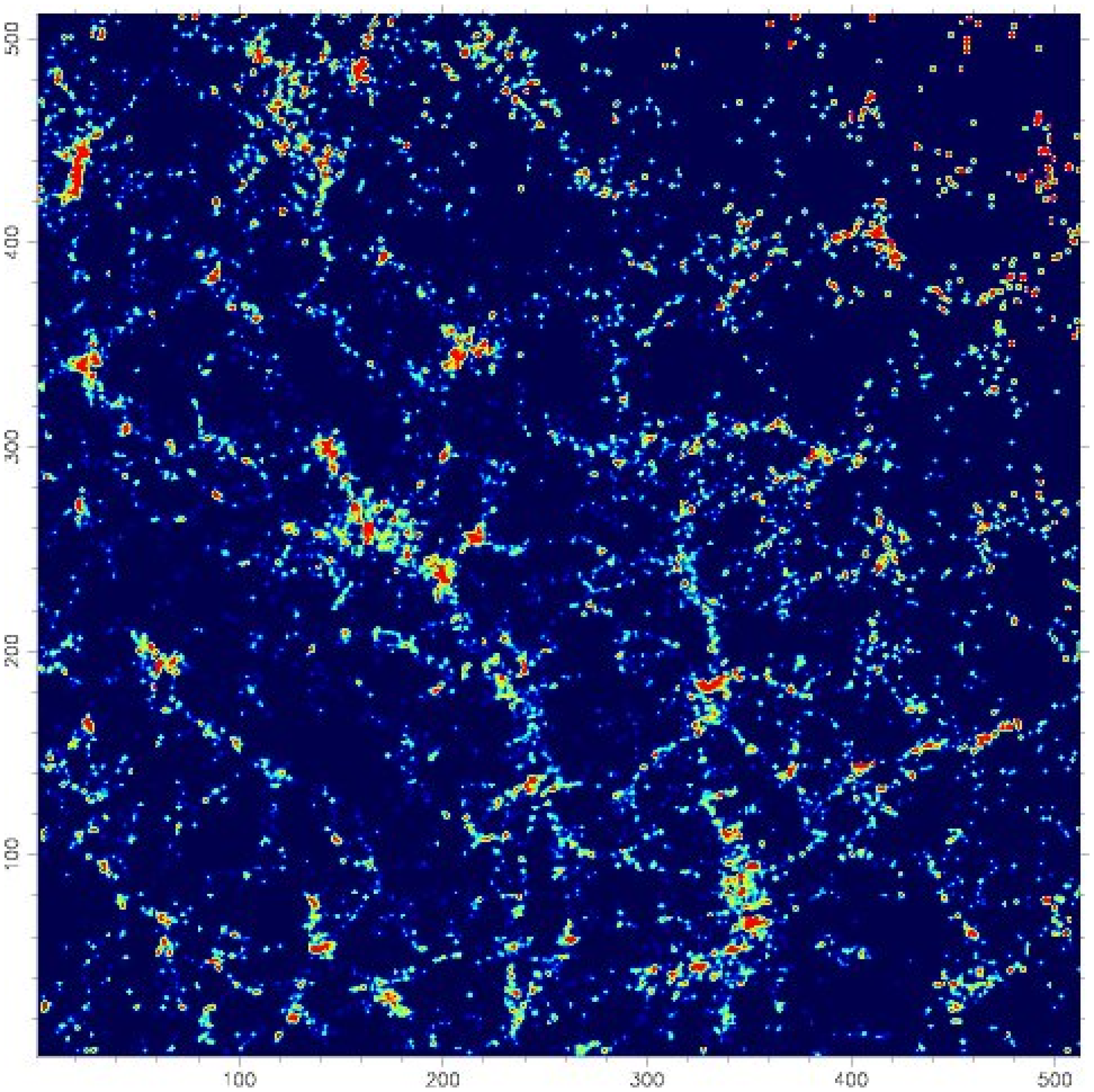}}
\hspace{2mm}
\resizebox{0.48\textwidth}{!}{\includegraphics*{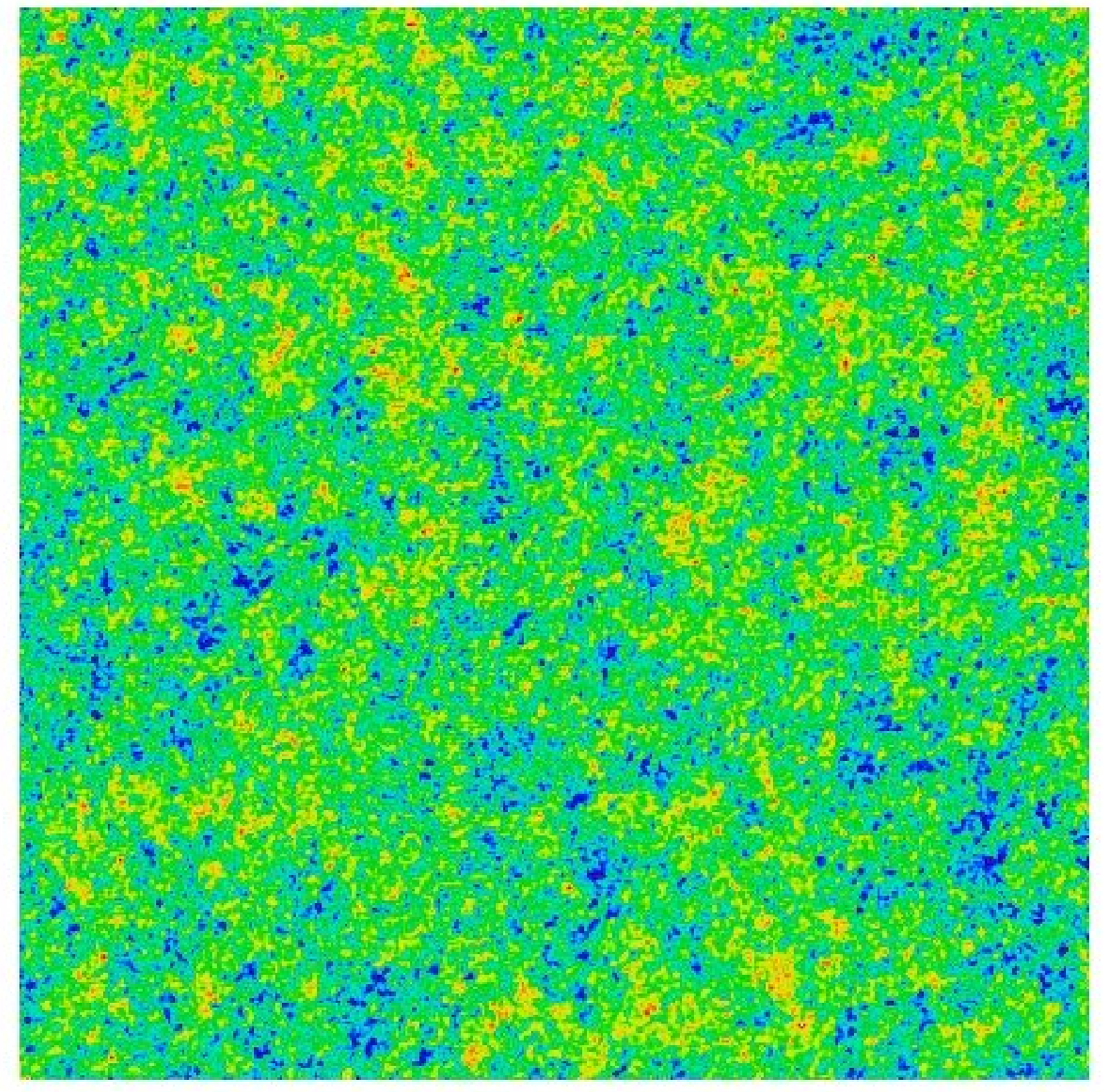}}\\
\resizebox{0.48\textwidth}{!}{\includegraphics*{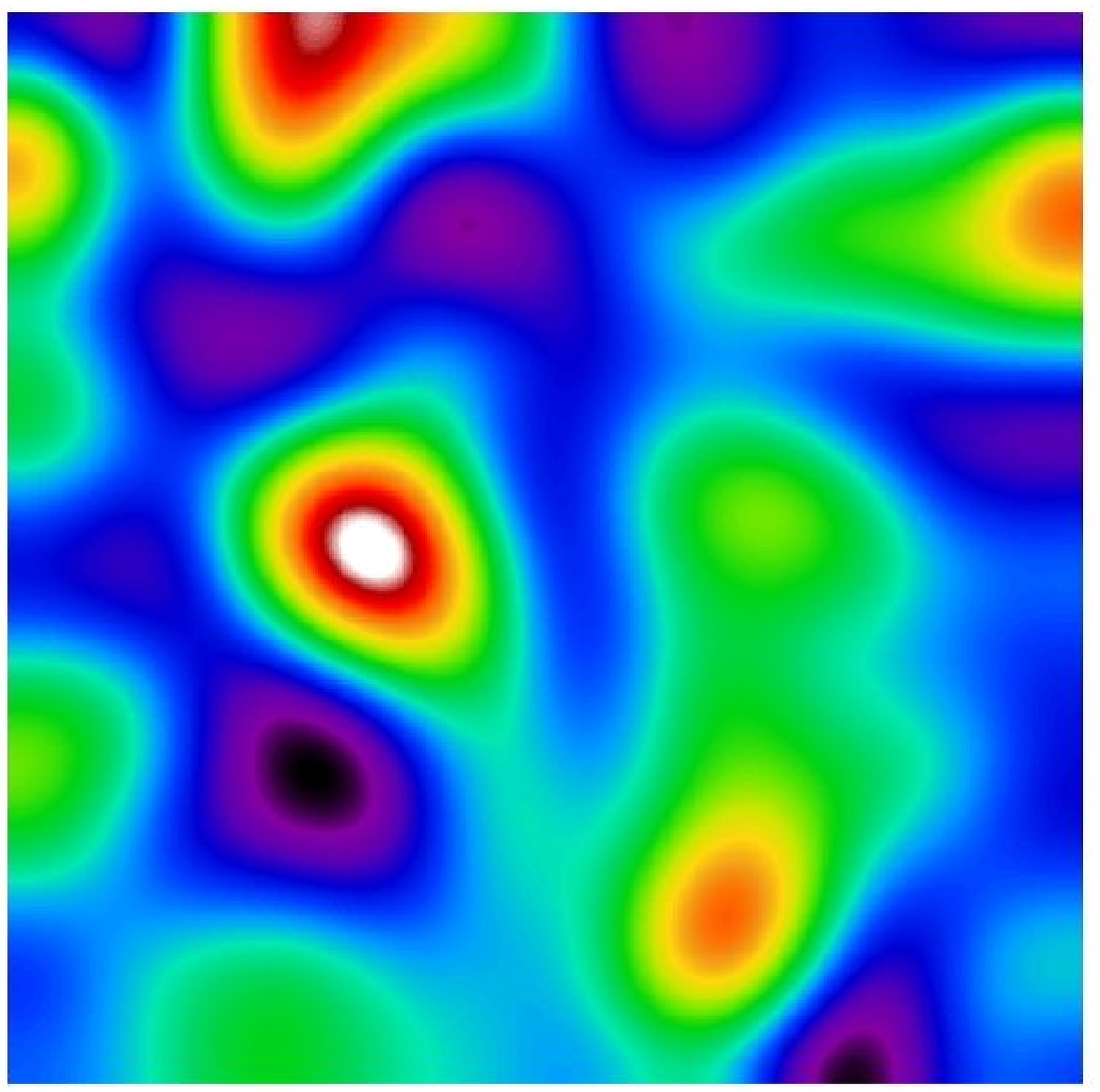}}
\resizebox{0.48\textwidth}{!}{\includegraphics*{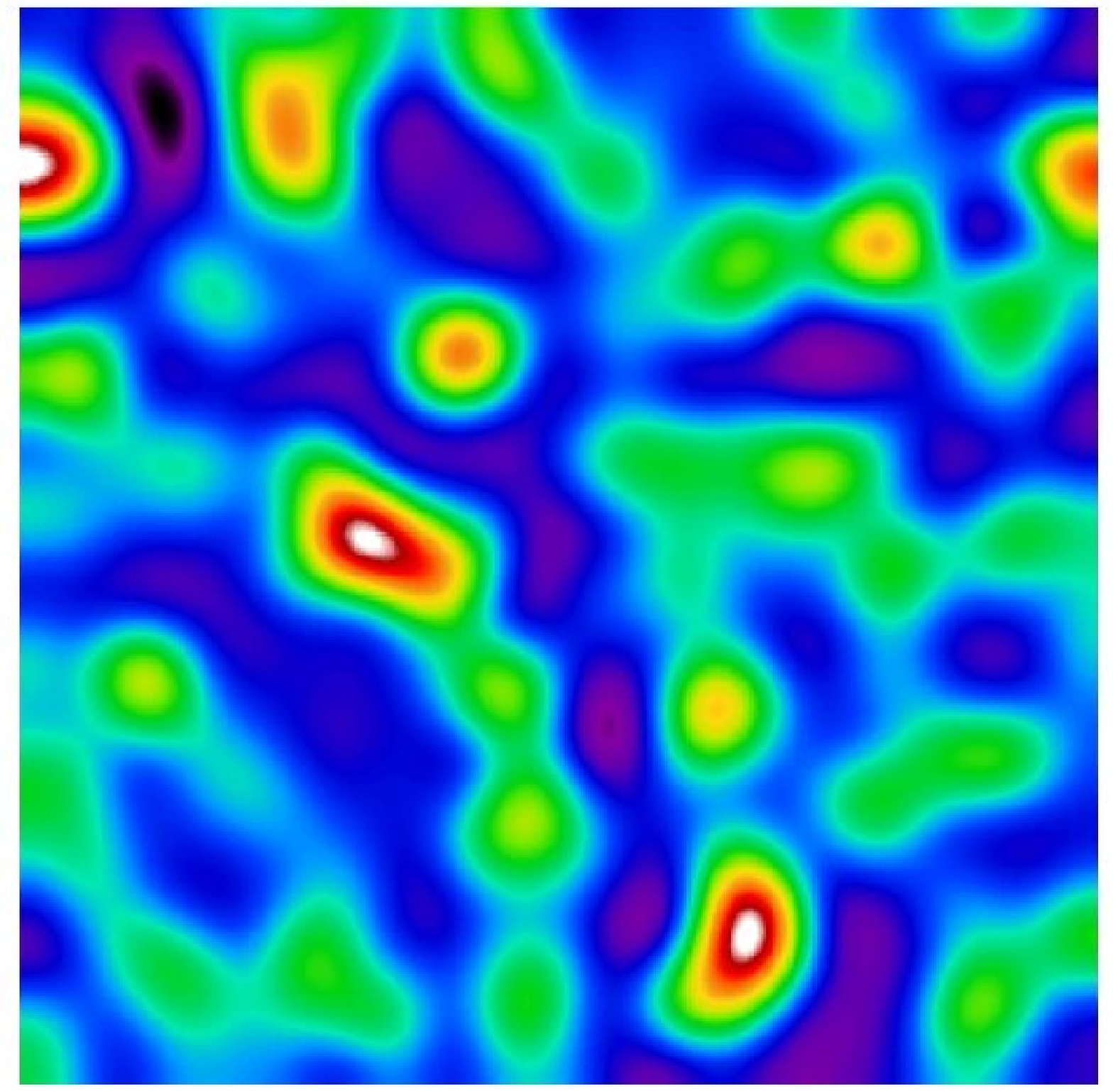}}\\
\resizebox{0.48\textwidth}{!}{\includegraphics*{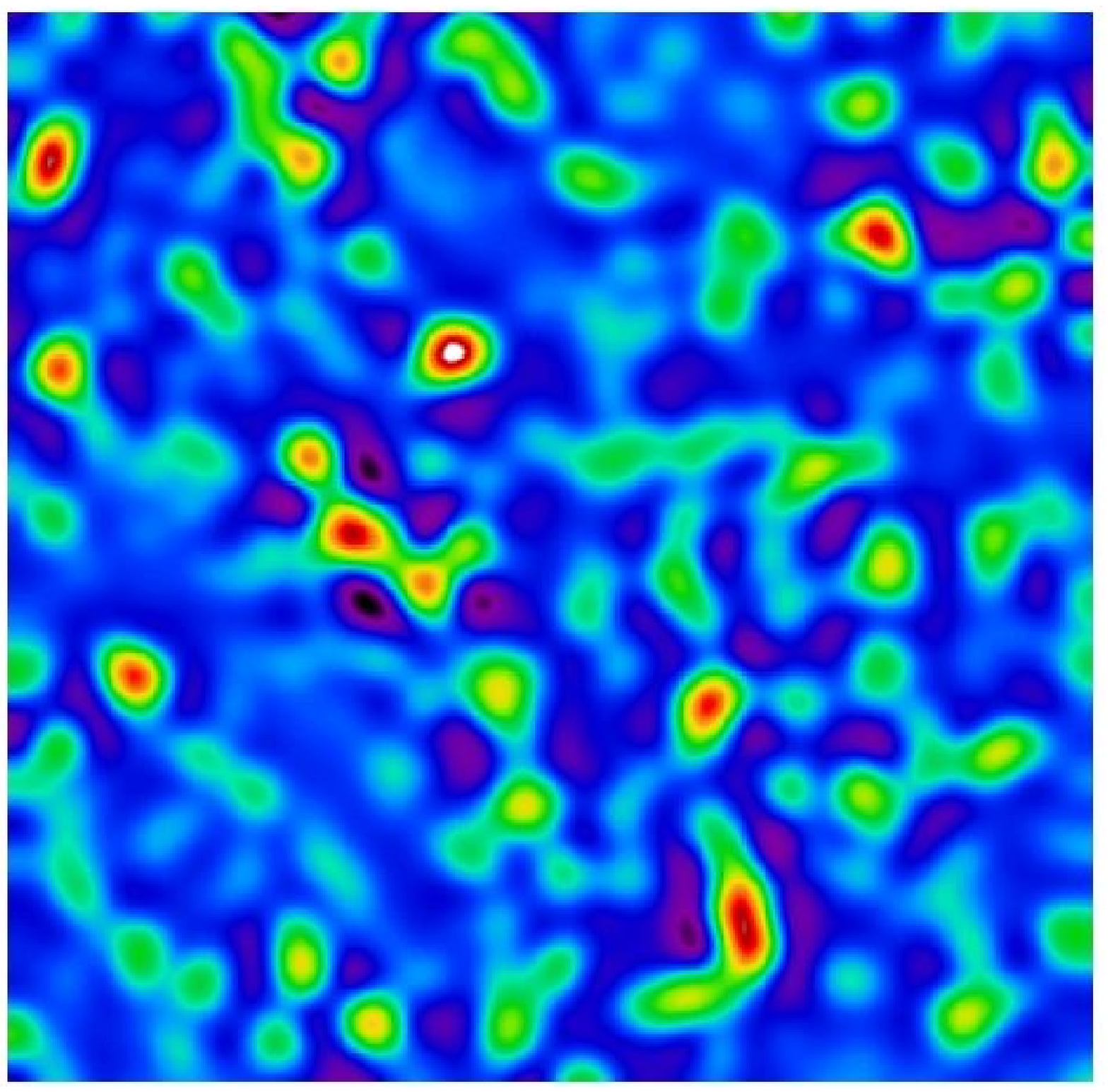}}
\resizebox{0.48\textwidth}{!}{\includegraphics*{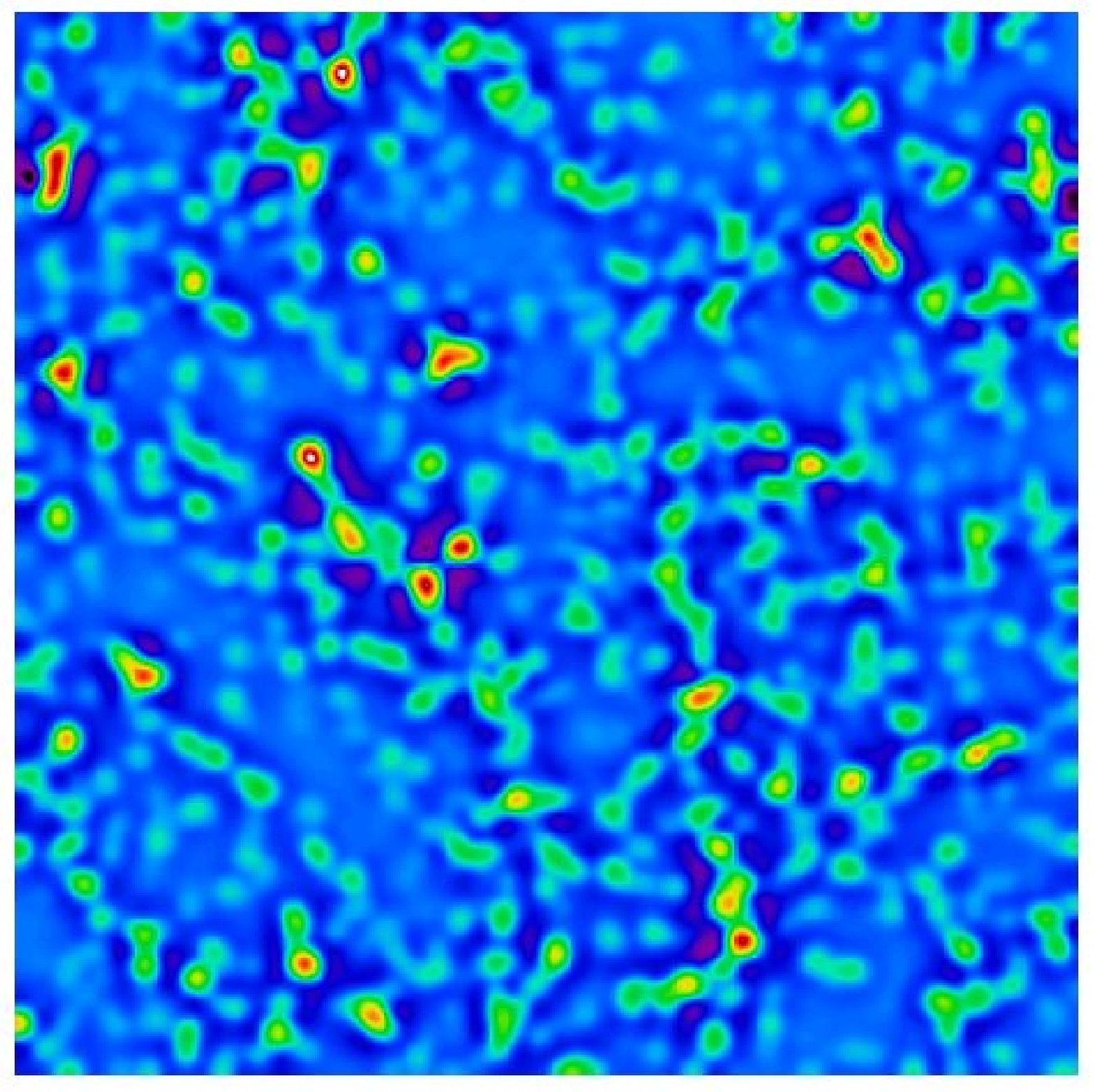}}
\end{minipage}
\caption{The upper left panel shows a rectangular region extracted from the
  high-resolution luminosity density field of the Northern equatorial slice of
  SDSS, and converted to distances using the dimensionless Hubble constant
  $h=0.8$. The density was calculated using Gaussian smoothing with a rms
  scale of 0.8~\Mpc.  The observer is located at the lower left corner.  The
  upper right panel shows the same density field as in the previous panel, but
  with randomly shifted phases.  The following panels show the wavelet
  analysis of the slice.  The middle left panel shows density waves of scales
  approximately equal to half of the box size, each following panel shows
  waves of scale 2 times smaller.  }
\label{fig:wavelet1}
\end{figure*}

As a first step we analysed a thin, almost 2-dimensional wedge of the Sloan
survey equatorial slice, 2.5 degrees thick, centered on the rich supercluster
SCL126 in the catalog by \citet{Einasto:2001} (the Sloan Great Wall). The
high-resolution luminosity density field, reduced to a constant thickness by
dividing the local density by the relative thickness of the wedge at a
particular distance, is shown in the upper left panel of 
Fig.~\ref{fig:wavelet1}.

The importance of phase information in the formation of the supercluster-void
network has been understood long ago, as demonstrated by Szalay (personal
communication) by randomizing phases of a Voronoi simulated picture of a
cellular network.  To study the role of phase information in more detail we
Fourier transformed the density field and randomized phases of all Fourier
components and thereafter Fourier transformed it back to see the resulting
density field which has for all waves the same amplitudes (power spectrum) as
the original field; see the upper right panel of Fig.~\ref{fig:wavelet1}.  We
see that the whole structure of superclusters, filaments and voids has gone,
the field is fully covered by tiny randomly spaced density enhancements --
groups of galaxies, there are even no rich clusters of galaxies in this
picture that were comparable in luminosity to the real clusters.

Next we used wavelet analysis to investigate the role of density waves
of different scale (for details of the wavelet analysis see
\citet{Martinez:2002}).  The full field was divided into 8 components,
so that the component with the largest waves represents waves of the
length about 512 Mpc (the size of the box), and each of the following
components represents waves of twice smaller scale. Actually, each
component contains waves of scale interval from $\sqrt{2}$ times lower
to $\sqrt{2}$ times higher than the mean wavelength, so that by
combining waves of all components we restore accurately the initial
density field.  Fig.~\ref{fig:wavelet1} shows components 4 to 7; they
characterize waves of length from about 32 to 256 Mpc.

Let us discuss the distribution of waves in more detail.  The middle left
panel of Fig.~\ref{fig:wavelet1} shows the waves of length about 256 Mpc.  The
regions of highest density of this component are three very rich
superclusters: located in the upper left part of the Figure, near the center,
and in the lower right part of the panel, the superclusters SCL155, SCL126 and
SCL82 in the list by \citet{Einasto:2001}, respectively.  The middle right
panel shows waves of scale of about 128 Mpc. Here the most prominent features are
the superclusters SCL126 and SCL82 as in the previous panel, but we see several
new superclusters: N15 and N04 in the list by \citet{Einasto:2003}.  The
supercluster N15 lies between SCL126 and SCL155 near the minimum of the 
density wave of 256 Mpc scale.

\begin{figure*}[ht]
\centering
\resizebox{0.48\textwidth}{!}{\includegraphics*{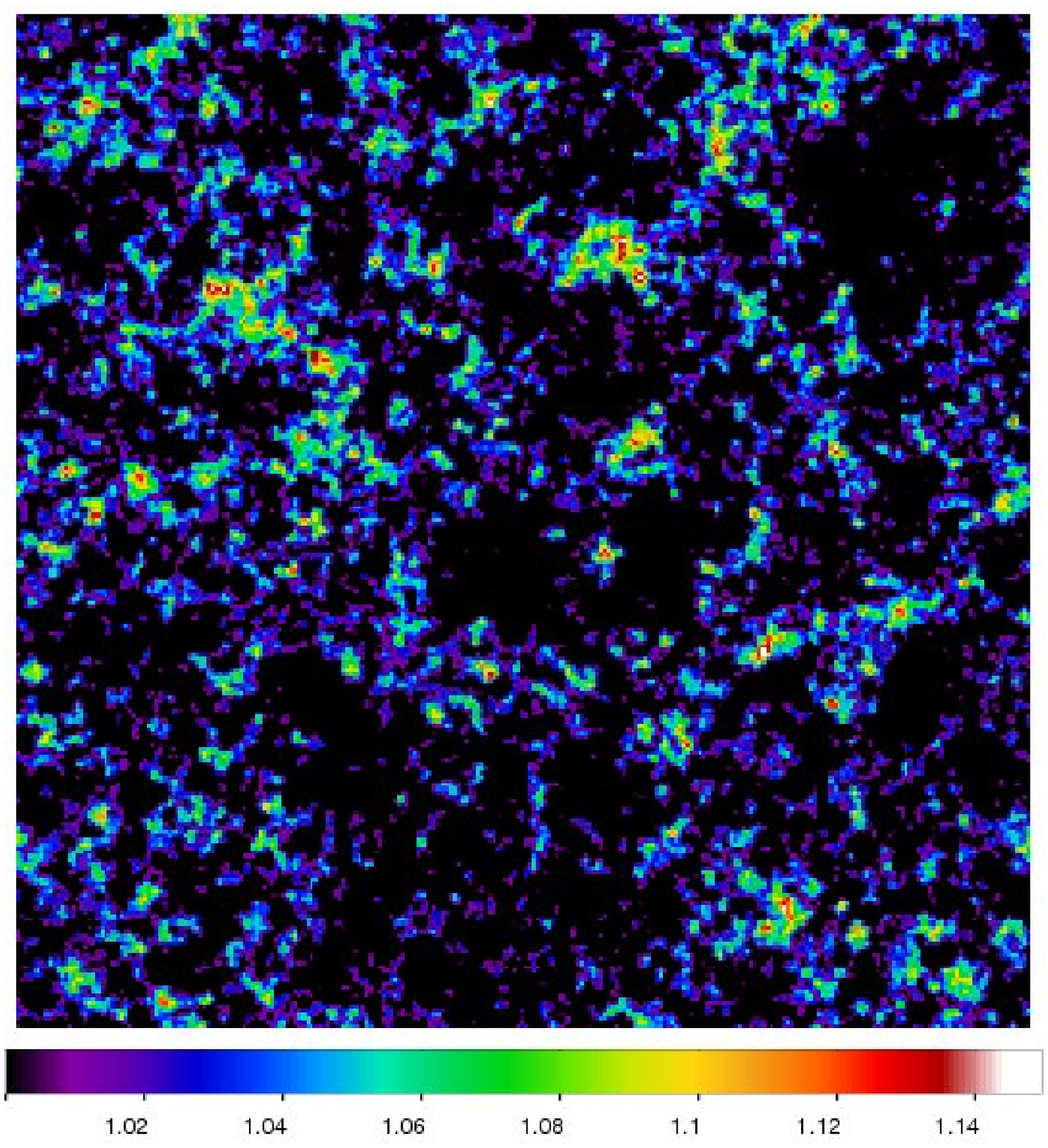}}
\resizebox{0.48\textwidth}{!}{\includegraphics*{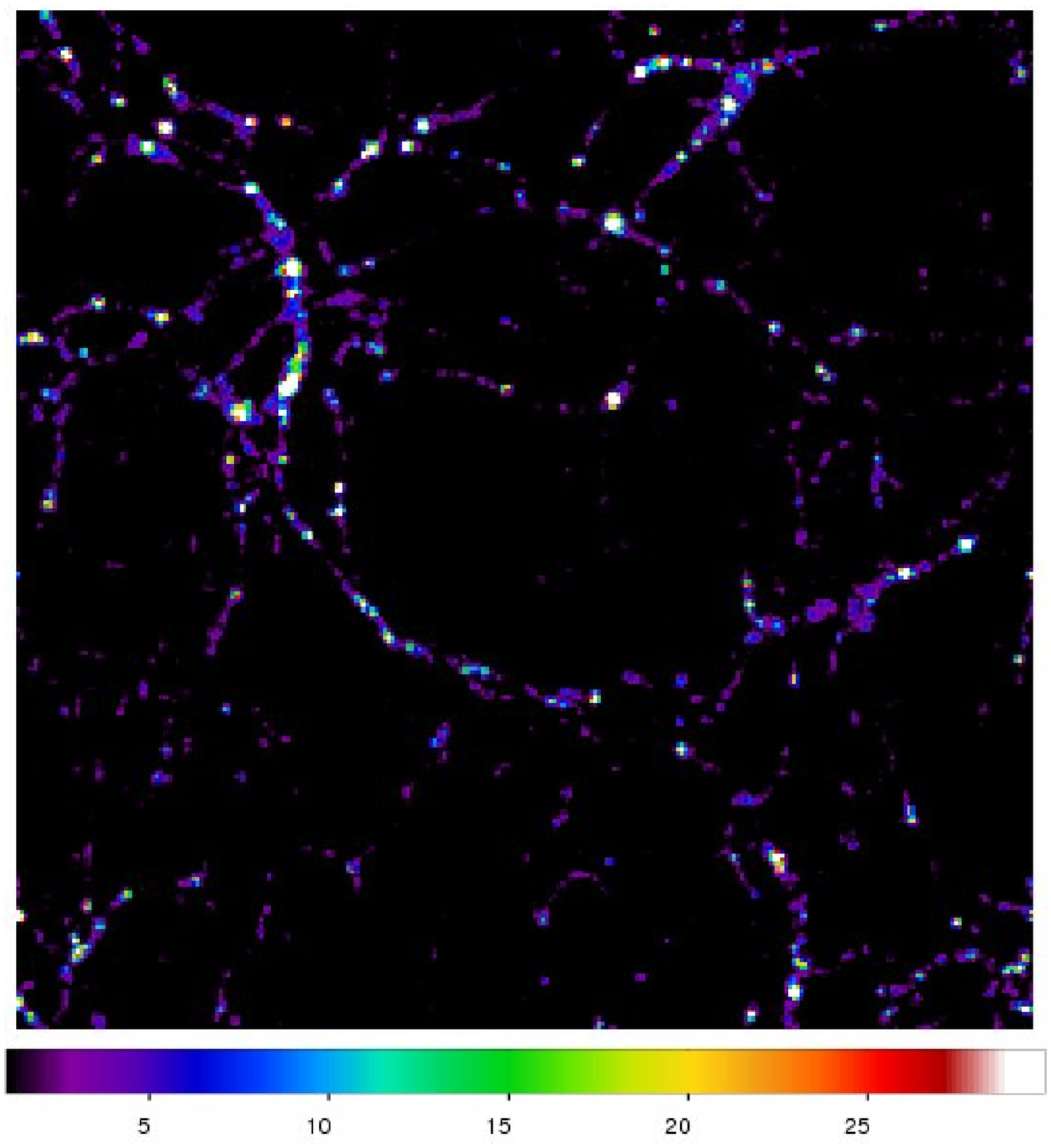}}
\caption{The high-resolution density field of the  model M100. Left
  panel shows the field at the redshift $z = 100$, the right panel at the present
  epoch ($z = 0$).  The color coding is chosen to show regions of overdensity,
  i.e. regions of local density greater than unity in mean density units.
}
\label{fig:evol}
\end{figure*}

The lower left panel plots waves of scale about 64 Mpc. Here all superclusters
that are seen at larger scales are also visible.  A large fraction of density
enhancements are either situated just in the middle of the low-density regions
of the previous panel, or they divide massive superclusters into smaller
subunits.  The highest peaks are substructures of massive superclusters, and
there are numerous smaller density enhancements between the peaks of the
previous panel.

This analysis allows to make several conclusions.  First of all, our results
show that in all cases superclusters form only in regions where {\em large
  density waves combine in similar phases of high-density regions to generate
  high density peaks}.  Very rich superclusters are objects where density
waves of all large scales (up to the wavelength $\sim 250$ ~Mpc) have similar
phases of density maxima.  The smaller the maximum wavelength of such phase
synchronization, the lower is the richness of superclusters.  Similarly, large
voids are caused by large-scale density perturbations of wavelength $\sim
100$~\Mpc, here large-wavelength modes combine {\em in similar phases of
  minima of density waves to generate under-densities}.

The second conclusion that can be made from the wavelet analysis
concerns the relation of waves of different scale.  Density
perturbations which have random phases and wavelengths in respect to
the principal perturbations (responsible to the formation of rich and
very rich superclusters) cancel partly each other out.  An exception
are the perturbations which have similar phases and wavenumbers that
are integer multiples of the wavenumber of the principal perturbation.
These are the harmonics (overtones) of the principal perturbation.
The first overtone has a wavenumber twice of the principal one; it
forms a density peak just in the middle of the void between two rich
superclusters, forming two sub-voids of approximately equal diameter
within the large under-dense region.  This behavior is well seen in
the wavelet decomposition of the density field in
Fig.~\ref{fig:wavelet1}.

The phase synchronization that generates rich superclusters is not surprising, if
considered from the point of view of the density field.  A high peak in the
field is possible only in the case if density waves of various scale up to a 
certain limiting wavelength have maxima at the same location.  The acoustic
behavior of density waves -- the domination of overtones -- deserves more
attention.  Presently it is not clear how to explain this behavior.

\section{The evolution of superclusters on simulations}

To investigate the role of density perturbations of different
wavelength we used N-body simulations with $256^3$ particles and with
box sizes $L=64,~100,~256,~768$~\Mpc.  Most models were calculated by
Ivan Suhhonenko in Tartu, the $L=100$~\Mpc\ model was obtained by
Volker M\"uller in Potsdam; we denote this model as M100.  In the
present review I shall analyse only the model M100, other models yield
similar results. In model M100 the following cosmological parameters
were used: the matter density $\Omega_m=0.27$, the baryonic matter
density $\Omega_b = 0.04$, the dark energy density $\Omega_\Lambda =
0.73$; the initial amplitude parameter $\sigma_8 = 0.84$, and the
dimensionless Hubble parameter $h=0.72$.  Calculations started at an
early epoch, $z=100$, the early evolution was calculated using the
Zeldovich approximation.  Particle positions and velocities were
extracted for the redshifts $z = 100, ~10,~5,~3,~2,~1,~0.0$. For every
particle we calculated also the local density in units of the mean
density, using positions of 27 nearby particles.

For all models and epochs we calculated the high-resolution density
field using the B3 spline smoothing with the step width equal to the
spatial resolution of the simulation (for details of smoothing with
the B3 spline see \citet{Einasto:2008}).  The high-resolution density
field of the M100 model at the epochs $z=100$ and $z=0$ is presented
in Fig.~\ref{fig:evol}.  Here we have selected from the full
simulation box a sheet that passes through a large void surrounded by
superclusters and filaments.  At the high-redshift epoch the density
contrast is much smaller as expected, but main features of the
filament system around the void are clearly seen.  The basic
high-density knots (protohalos of present-day rich clusters) are very
well seen already at the early epoch, and their positions have changed
only very little.

\begin{figure}[ht]
\begin{minipage}[h]{0.48\textwidth}
\centering
\resizebox{0.48\textwidth}{!}{\includegraphics*{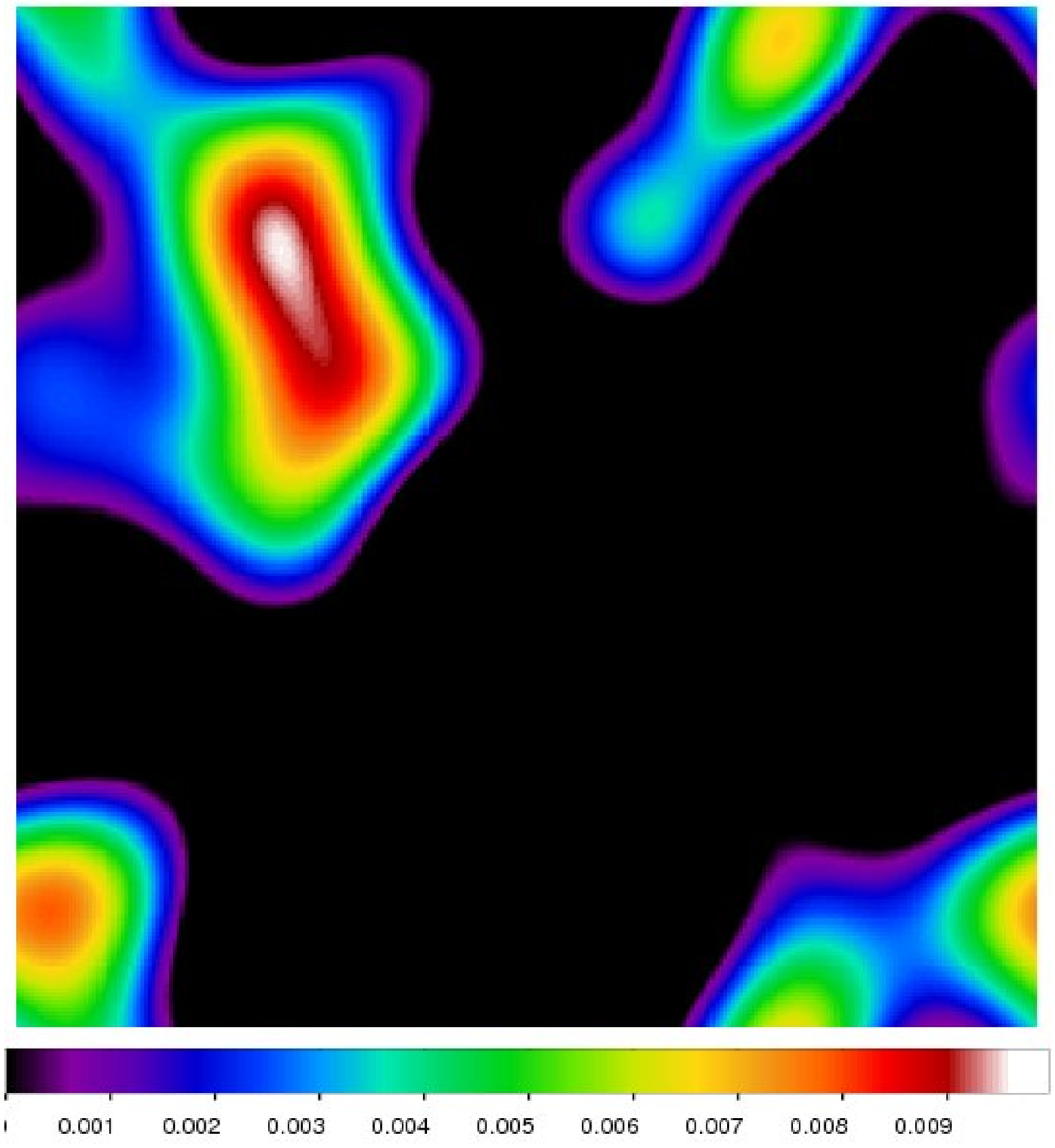}}
\resizebox{0.48\textwidth}{!}{\includegraphics*{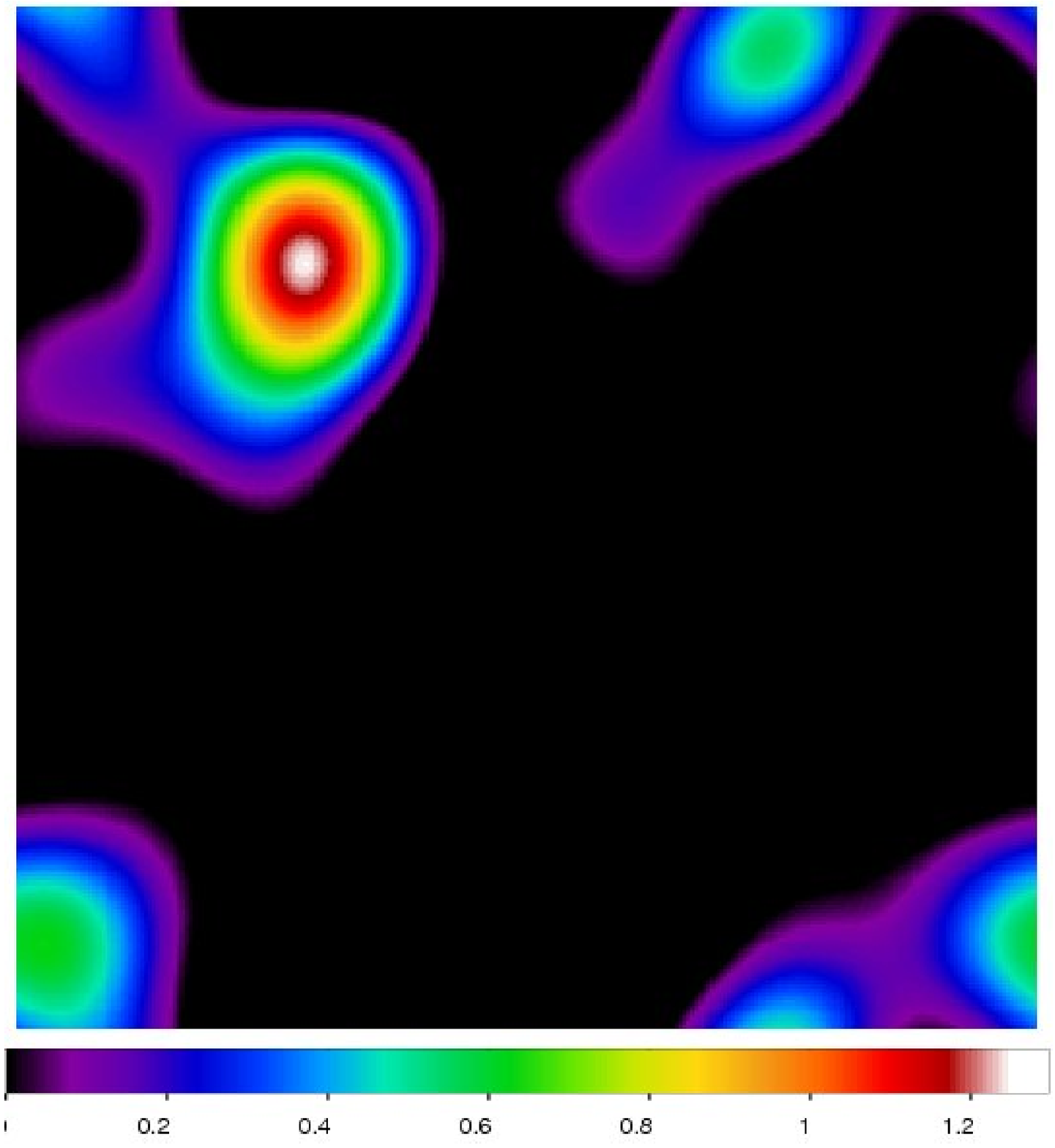}}\\
\resizebox{0.48\textwidth}{!}{\includegraphics*{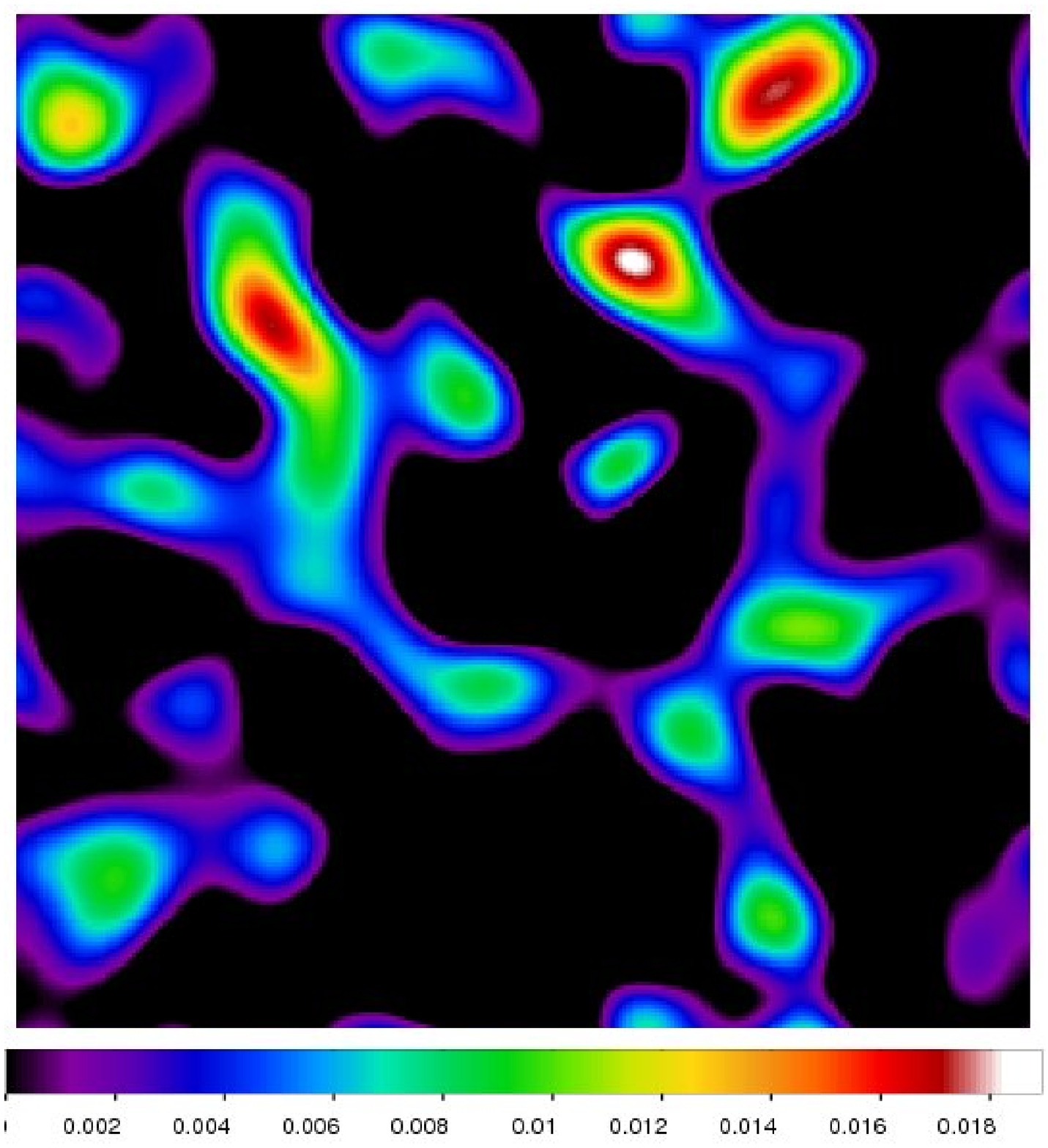}}
\resizebox{0.48\textwidth}{!}{\includegraphics*{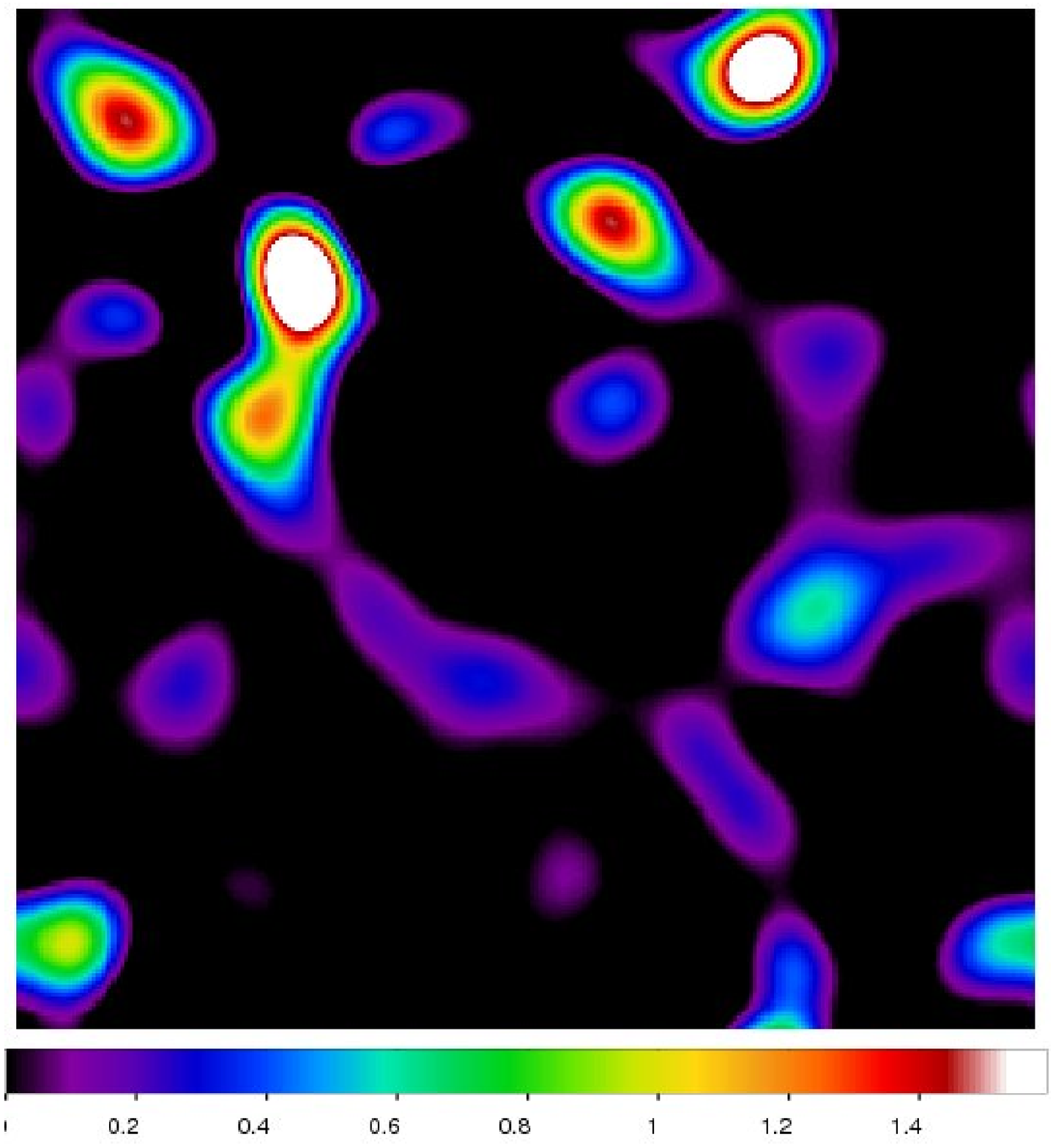}}\\
\resizebox{0.48\textwidth}{!}{\includegraphics*{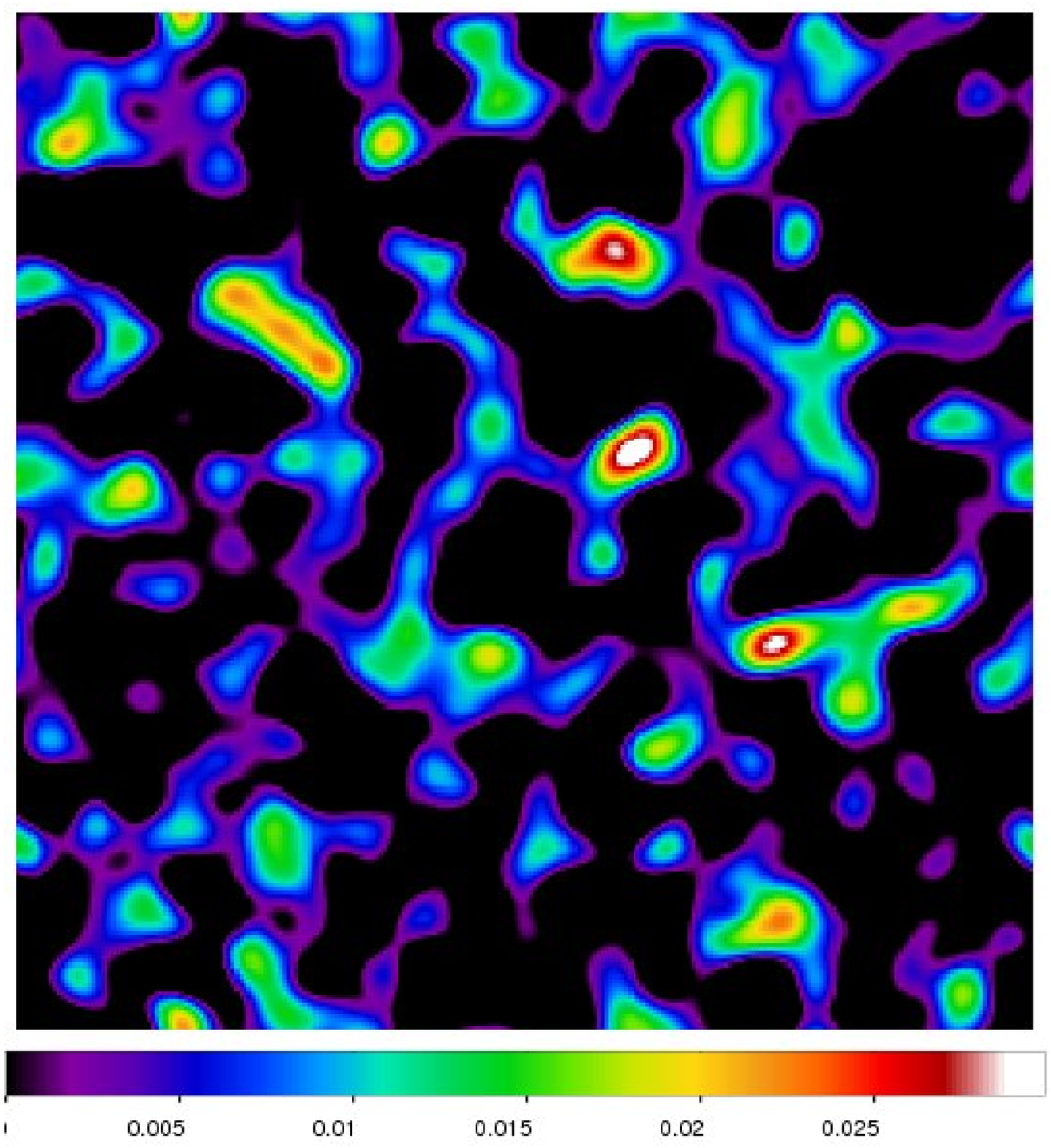}}
\resizebox{0.48\textwidth}{!}{\includegraphics*{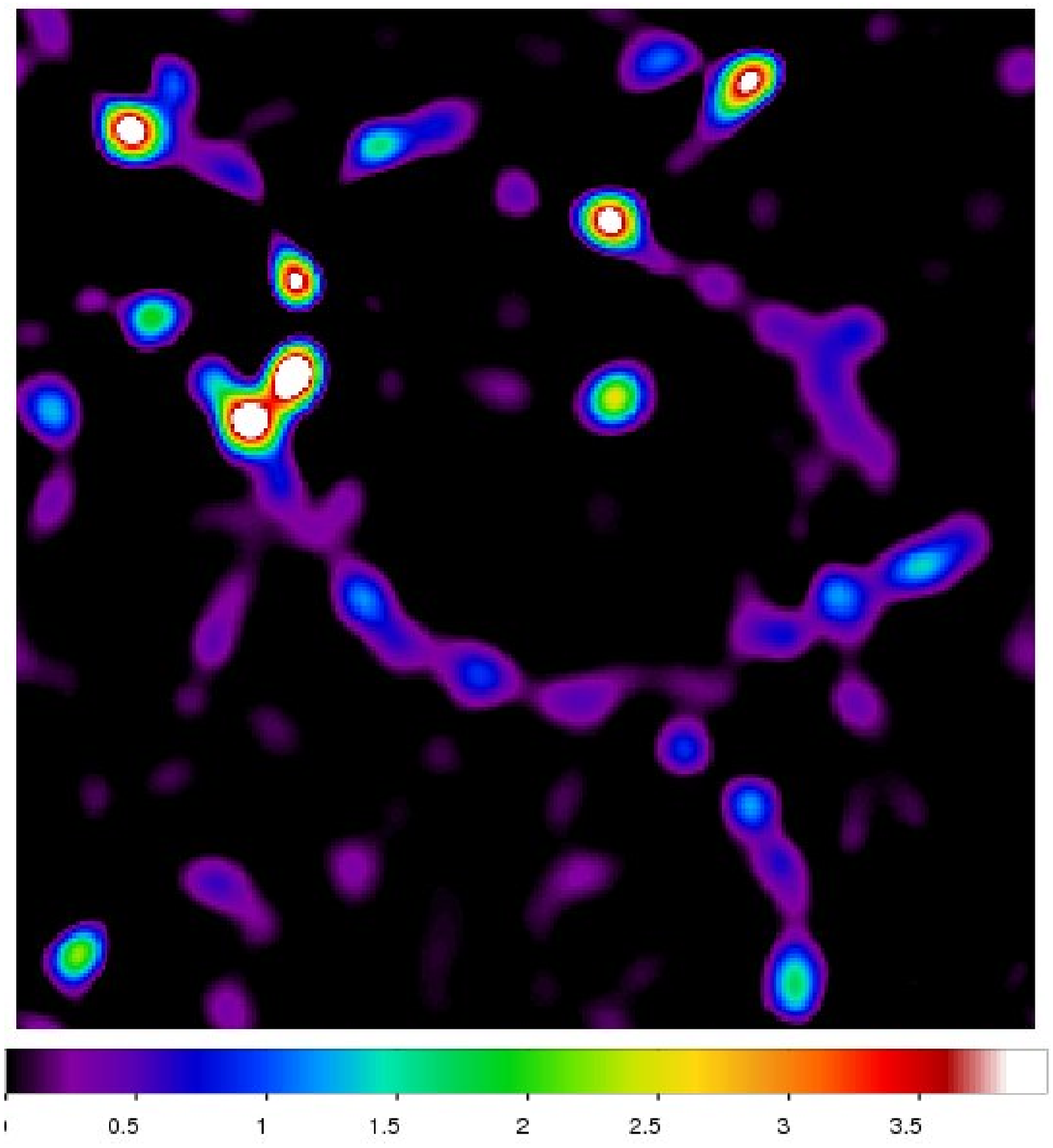}}\\
\end{minipage}
\caption{Wavelet analysis of the model M100 shown in
  Fig.~\ref{fig:evol}.  Left panels show density waves of various
  scales at the redshift $z = 100$, right panels at the present epoch
  ($z = 0$). In the upper panels the characteristic wavelength is
  approximately equal to the size of the box, $L = 100$~\Mpc, in the
  middle panels the scales are twice smalle, and in the lower panels
  four times smaller.  }
\label{fig:wavelet2}
\end{figure}

The Figure~\ref{fig:wavelet2} shows the wavelet decomposition of the same
field.  Left panels show the high-redshift, and right panels the present epoch
decomposition.  Wavelets of the order from 6 (upper panels) to 4 (lower
panels) are shown, they show waves from the size of the box (100~\Mpc) to about
25~\Mpc.  Only overdensity regions are shown, having densities above unity in
mean density units.  The colour code is chosen so that it corresponds
approximately to the factor of linear growth from $z=100$ to $z=0$,  about
76.
 
The Figure shows that large-scale waves have almost the same shape in the
early and present epoch, only the amplitude has grown as expected.  Even the waves
of scale 25~\Mpc\ have changed only little -- the main features at both epochs are
the same, only some filaments have joined during the evolution.  This is the
well-known effect of structure merging inside superclusters. 

It is well known that the Zeldovich approximation of linear growth is the more
accurate the earlier epochs and larger waves we consider.  Thus it is evident
that at epochs earlier than our first epoch discussed here, the growth can be
described very accurately by the Zeldovich approximation.  In other words:
all large-scale structures have had almost no evolution in their positions of
density wave features (maxima, minima and other similar features), the
evolution of these features is only in the amplitude of the density contrast.

The second important lesson from this analysis concerns the positions of
maxima of waves of various scale.  When we take a density maximum of the wave
of the largest scale, then at the same position ALL waves of smaller scale
have also a maximum, i.e. maxima of density waves are synchronized.  The same
behavior is present in the real density field of Sloan galaxies as discussed
above. 

Thus wavelet analysis of our simulations demonstrates that positions
of high-density objects (superclusters, rich clusters) almost do not
change during the whole history of the evolution of structure.  In
other words: the skeleton of the cosmic web or the supercluster-void
network was fixed already at a very early stage, much before the epoch
of recombination.  This conclusion confirms early analytic
calculations by \citet{Kofman:1988}, where main contours of the cosmic
web were found using the theory of adhesion at the epoch which
corresponds to the end of the inflation stage and to the start of the
Hubble expansion.

Our analysis shows that high-density peaks of the density field are present
already at the very early stage of the evolution of the structure.  Recent
numerical studies show that the formation of first generation stars starts
just inside these high-density peaks.  The density is highest in protohalos of
central clusters of future rich superclusters.  Simulations by
\citet{Gao:2005b} have shown that metal-free gas in dark matter haloes of
virial temperature about 2000 K and mass $M \sim 10^6$~M$_\odot$ cools
efficiently, thus giving rise to formation of stars.  In these high-density
regions star formation can start as early as at the redshift $z\approx 50$.
These first generation stars have large masses and evolve rapidly, and at
least some of them may explode and spread products of nuclear synthesis to the
surrounding ``pure'' gas.

\section{Conclusions}

\begin{itemize}

\item{}Superclusters form in regions where large-scale density waves combine in
  similar locations of maxima.  

\item{}Superclusters are the richer the larger is the wavelength of wave
  synchronization. 

\item{}Medium rich superclusters are often formed near 	 minima of waves of
  scale $\sim 250$~\Mpc\ by the first overtone of this wave. 

\item{}Voids form in regions where large-scale density waves combine in
  similar locations of minima.   

\item{}Large-scale density waves do not change positions of maxima during the
  evolution,   i.e the web of the filaments that defines the final state of the
  large-scale galaxy distribution is actually present in the initial density
  inhomogeneities.

\item{}In centers of rich superclusters all waves of various scale
  have maxima. Since positions of large waves do not change, high
  local densities were formed already at a high redshift $z \geq
  100$. Thus formation of the first generation stars started in the
  centers of rich superclusters.

\end{itemize}

\begin{theacknowledgments}
  I thank Enn Saar, Ivan Suhhonenko, Lauri-Juhan Liivam\~agi, Elmo
  Tempel, Maret Einasto, Erik Tago and Volker M\"uller for
  collaboration and for the permission to use our common results in
  this review talk.  The present study was supported by the Estonian
  Science Foundation grant ETF 8005, and by the grant SF0060067s08.  I
  thank the Astrophysikalisches Institut Potsdam (DFG-grant MU
  1020/11-1 436), ICRAnet, and the Aspen Center for Physics, where
  part of this study was performed, for hospitality.
\end{theacknowledgments}

{\small

}

\end{document}